# Local tissue effects and peri-implant bone healing induced by implant surface treatment: an *in vivo* study in the sheep

Nicolas Rousseau[1,2,*], Inès Msolli[2], Patrick Chabrand[1], Arnaud Destainville[3], Olivier Richart[2], and Jean-Louis Milan[1]

[1]*Aix Marseille Univ, CNRS, ISM, Marseille, France*
[2]*Selenium Medical, 9049 rue de Québec, CS80875, 17043 La Rochelle, France*
[3]*Abys Medical, 40 rue Chef de Baie, 17000 La Rochelle, France*

[*]Corresponding author: Nicolas Rousseau, Selenium Medical, 9049 rue de Québec, CS80875, 17043 La Rochelle, France. Tel: +33 5 46 44 40 82; Fax: +33 5 46 31 62 71; Email: nrousseau@selenium-medical.com.

**Running title:** Tissue effects and bone healing induced by topography

## Acknowledgements

The authors would like to thank Namsa (Chasse-sur-Rhône, France) for providing the animal model, housing and husbandry and performing the different surgical procedures, histopathologic and histomorphometric analyses. The authors also acknowledge Generic Implants (Saint-Genis-Laval, France) for supplying the implants used for the study and Selenium Medical (La Rochelle, France) for the surface treatment and packaging applied to these implants.

## Conflict of Interest Statement

The study was financially supported by the company Selenium Medical. Some authors are employees of Selenium Medical and took part in the study design, implants physicochemical characterization and the writing of this article. However, the authors certify that they were not involved in surgical procedures nor in the micro-computed tomography, histopathologic and histomorphometric analyses, which were conducted by an external expertise.

## Authors contributions

# Abstract

**Objective:** The aim of this study was to assess, through biological analysis, the local effects and osseointegration of dental implants incorporating surface micro/nanofeatures compared to implants of identical design without surface treatment. **Background**: Known to impact bone cell behavior, surface chemical and topography modifications target improved osseointegration and long-term success of dental implants. Very few studies assess the performance of implants presenting both micro- and nanofeatures *in vivo* on the animal models used in preclinical studies for medical device certification. **Methods:** Implant surfaces were characterized in terms of topography and surface chemical composition. After 4 weeks and 13 weeks of implantation in sheep femoral condyles, forty implants were evaluated through micro-computed tomography, histopathologic, and histomorphometric analyses. **Results:** No local adverse effects were observed around implants. Histomorphometric analyses showed significantly higher bone-to-implant contact in the coronal region of the surface treated implant at week 4 and week 13, respectively 79.3±11.2% and 86.4±6.7%, compared to the untreated implant's 68.3±8.8% and 74.8±13%. Micro-computed tomography analyses revealed that healing patterns differed between coronal and apical regions, with higher coronal bone-to-implant contact at week 13. Histopathologic results showed, at week 13, bone healing around the surface treated implant with undistinguishable defect margins while the untreated implant still presented bone condensation and traces of the initial drill defect. **Conclusion:** Our results suggest that the surface treated implant not only shows no deleterious effects on local tissues but also promotes faster bone healing around the implant. (word count: 241)

# 1 Introduction

Dental implants are commonly used for the oral rehabilitation of partially or fully edentulous patients. To improve osseointegration and ensure that implants remain well-integrated in bone long-term, researchers and manufacturers continue to develop new implant designs, altering their macrogeometry and modifying their surface at lower scales. Among the many surface modifications applied to improve initial and long-term osseointegration, the sandblasted acid-etched (SLA) surface treatment has been for several decades one of the most popular surface treatment on the market. This surface treatment provides dental implants with a microrough texture that improves bone apposition and biomechanical anchorage (1). Plasma spraying has also been used to obtain microrough titanium surfaces (2) or to generate a hydroxyapatite coating on the implant surface (3). However, clinical concerns were raised about reported hydroxyapatite coating resorption (4) and the long-term efficiency of plasma spraying was questioned (5). Current research focuses particularly on the combination of micro- and nanofeatures obtained, for instance, by acid-etching followed by oxidation in flowing synthetic air (6), or by anodization yielding titanium nanotubes at the implant surface (7).

Generally speaking, it is accepted by the scientific literature that modifying the surface of implants in terms of chemical composition and topology affects their rate of osseointegration. In vivo studies demonstrated that micrometric-scale features introduced on titanium surfaces enhance bone-to-implant contact (BIC) 8 weeks following implantation, compared to smoother surfaces (8,9).

In vitro studies confirmed such observations, reporting lower numbers of osteoblasts at the interface between bone and rough surfaces, along with increased differentiation markers such as alkaline phosphatase and osteocalcin (10,11). The authors of these studies suggested that osteoblasts are sensitive to moderate micro-roughness, moving earlier from their proliferation phase to their differentiation phase, maturing faster, and rapidly starting to secrete extracellular matrix (ECM). On the other hand, surfaces presenting nanofeatures showed improved BIC and higher implant removal torque in vivo two weeks after implantation (12), as well as better adhesion of osteoblasts in vitro (13). It should be noted that combining microfeatures and nanofeatures generates additive effects at both topographic levels, in addition to a drastic increase in surface hydrophilicity (14,15). To our knowledge, very few in vivo studies assessed the combined influence of micro and nanofeatures on dental implant osseointegration, and these were only conducted either on small animals (16,17) or for very short periods (18). Recently, Liu et al. (2019) studied screw-shaped cylinders implanted for three months in osteoporotic sheep mandibulars and found higher biomechanical parameter values for surfaces presenting micro/nanofeatures (19). However, no assessment of the neo-bone formation, of fibrosis, nor of the bone texture was conducted in this study. Furthermore, none of these studies made a distinction between the coronal area and the apical area of the implant, and thus no difference in osseointegration could be observed. Distinguishing between the two areas would have enabled the local influence of implant design features to be assessed.

The purpose of this paper is twofold: (1) to assess the performance (osseointegration and osteogenesis) of a dental titanium implant presenting micro/nanofeatures; (2) to provide a comprehensive biological analysis of the local tissue effects of such an implant. Sheep were implanted with dental implants of a unique design, both untreated and treated with Starsurf® (Selenium Medical, La Rochelle, France), a chemical surface treatment that provides micro/nanofeatures. A total of eight sheep were bilaterally implanted in the medial femoral condyle and sacrificed four weeks and thirteen weeks after implantation. In addition to surface topography characterization and chemical composition assessment, Bone-to-Implant Contact (BIC) was evaluated through histomorphometric analysis to address the primary study objective both in the coronal and the apical regions. Local tissue effects of the endosseous implants were evaluated qualitatively and semi-quantitatively through histopathologic analysis. Finally, micro-computed tomography analyses were conducted to provide supporting data for both types of analysis.

## 2 Materials and method

### 2.1 Ethical Statement

The implantation protocol was approved by the Namsa Ethical Committee (Chasse-sur-Rhône, France) as part of a project authorization reviewed every five years by the French Ministry of Education, Higher Education and Research. The different surgical procedures and biological analyses were conducted by the Medical Research Organization Namsa (Chasse-sur-Rhône, France), in compliance with the OECD series on Good Laboratory Practice and compliance monitoring and FDA 21 CFR 58 on Good Laboratory Practice (GLP) for non-clinical studies. The study design was also based on the guidance provided in ISO 10993-6, Biological Evaluation of Medical Devices Part 6 (2016): Tests for Local Effects after implantation. The animal study was conducted in accordance with the ARRIVE guidelines (20). The relevance and use of animal selection were carefully established and considered.

## 2.2 Sample size calculation

The sample size calculation for this study was based on a power analysis (21), considering a level of significance at 5% ($P$=0.05), a study power of 80% and assuming two-tailed hypothesis tests. Based on scientific literature (1), an effect size of at least 23% with a maximal standard deviation of 14% were expected for the BIC in this study. According to the calculation, a sample size of at least six implants per group is required. To obtain a finer statistical analysis and to avoid any adverse effect and unexpected event, ten implants were allocated per group as well as two reserve implants per group. Therefore, to study two types of implants at two time periods, forty-eight implants were necessary. With up to three screw-type implants per leg, eight sheep were required for the study. Such calculation was concomitant with qualitative and semi-quantitative evaluation of local tissue effect, as any level of difference was of interest in the absence of similar data in the literature.

## 2.3 The implants

Forty-eight implants commercially available and made of commercially pure titanium (cpTi, grade IV) were used (Fig. 1), including eight reserves similarly treated and only intended for replacement in case of an adverse event. The intra-osseous part of the implant has a diameter of 3.5 mm and a length of 10 mm, one part being cylindrical (coronal area) and the other conical (apical area). The main thread is composed of a double fillet with a pitch of 2 mm. The conical part of the implant has a thread depth that increases from approximately 0.15 mm in the coronal area to 0.7 mm in the apical one. In the coronal area, the main thread is subdivided into two "micro" fillets of identical pitch. Such implant geometry is common and representative of implants currently on the market (22). Two configurations were tested in this study: a standard topography (REF, n=10 + 2 reserves at both 4 weeks and 13 weeks) and a Starsurf® one (TEST, n=10 + 2 at both 4 weeks and 13 weeks) combining micro/nanofeatures. The standard topography is obtained through corundum grit-blasting, while the Starsurf® topography results from a succession of mechanical and chemical treatments. First, implants are sandblasted using a mixture of hydroxyapatite (HA) and $\beta$-tricalcium phosphate ($\beta$-TCP), a biocompatible material and a bone mineral constituent. In addition to the surface roughness and particular topography this grit-blasting process provides (23,24), hydroxyapatite is fully soluble in acid solutions. Therefore, subsequent acid attack by hydrochloric and sulfuric acid dissolves HA particles along with surface etching (25). Finally, a basic attack is conducted using potassium hydroxide. Both series of implants were cleaned in clean rooms, and sterile packaging and final gamma irradiation were applied prior to implantation. Following implantation, the threads were engaged in the host bone cylindrical wall obtained after drilling, providing primary mechanical stability.

## 2.4 Implant surface characterization

Under a Captair Flow 468 ISO5 laminar flow hood (AirLab), the overall topography of REF and TEST implants was qualitatively examined using a Phenom ProX Scanning Electron Microscope (ThermoFisher scientific). Micrographs at different magnifications were observed to assess the topographies at different scales. Then, surface chemical composition was assessed by Energy Dispersive X-Ray Spectrometry (EDX). Next, roughness measurements were conducted on an Altisurf 3D profilometer (Altimet). Areas of 200×750$\mu$m for the interspaces of threads were evaluated in both coronal and apical areas. The arithmetic mean deviation (Sa), root-mean-square deviation (Sq), maximum peak-to-valley height (Sz) and developed interfacial area ratio (Sdr) of the surface were extracted, constituting the main roughness parameters. Three REF and TEST implants were examined and measurements were performed in triplicate. Finally, REF and TEST surfaces were reproduced on cpTi discs to assess implant surface energy. Wettability was tested through contact angle measurements on a DSA25 tensiometer (Krüss). Surface free energy was assessed using the Owens-Rankine formulation, with three liquids, droplets of 2$\mu$l, and ten contact angle measurements per liquid.

## 2.5 Study design and experimental animals

**Animal model and management**

Eight (n= 4 per time period) female Blanche du Massif Central sheep (Bergerie de la Combe aux Loups, France) were involved in the study. Sheep were aged from 2 to 4 years (mean=2.7 years), skeletally mature, and weighed from 57 to 68 kg at implantation (mean=61.25 kg). The sheep is an animal model identified as suitable for evaluating materials and is suggested in the ISO 10993 standard - part 6 (2016) for intraosseous implantations. Sheep were bilaterally implanted in the medial femoral condyle. Sheep femurs can be implanted with up to three screw-type implants per leg, increasing the number of sites per implant group without increasing the number of animals. Sheep were randomly attributed to the two periods of implantation studied and the number of sites per implant group was chosen in accordance with the guidance provided in ISO 10993 standard - part6. One group per condyle was implanted to allow suitable evaluation of the local tissue effects and performance and to avoid mixed tissue response between implant groups. Husbandry, housing, and environment conditions were in conformity with European Directive 2010/63/EU regarding the protection of animals used for scientific purposes. Animals were housed at Namsa, an AAALAC international accredited facility registered with the French Department of Agriculture for animal housing, care, and investigations. They were grouped in cages identified by a card indicating the study number, number of animals, sex, dates of beginning and end of experimental in-life phase. The animal housing room temperature and relative humidity were recorded daily. The recommended temperature range for the room was 10 - 24°C and the light cycle was controlled using an automatic timer (12 hours of light, 12 hours of dark). After the post-operative period, the sheep from the 13-week group were jointly housed in a farm setting (Bergerie de la Combe aux Loups, France), where environmental conditions were not controlled. Standard hay was provided *ad libitum* and supplemented with a commercially available pelleted sheep feed (Special Diet Services, France). Minerals were provided *ad libitum* (Sodimouton, Salins Agriculture). Drinking water was delivered *ad libitum* through species-appropriate containers or delivered through an automatic watering system.

**Animal selection and randomization**

Only healthy and previously unused animals were selected for this study. Each animal was randomly assigned to one of the two time periods. Three implants of the same configuration were installed per femoral condyle and each animal was implanted with both implant configurations. The implantation site of each implant was randomly allocated. The randomization was carried out electronically by an independent author involved neither in the animal selection nor in the surgical procedures.

**Pre-operative procedure**

On the day of surgery, pre-medication was performed by intravenous injection of a mixture of diazepam (Diazepam®, TVM) and butorphanol (Torphasol®, Axience). Anesthesia was induced by intravenous injection of propofol (Propovet®, Zoetis). Each sheep was intubated, mechanically ventilated, and placed on isoflurane inhalant anesthetic (IsoFlo®, Zoetis) for continued general anesthesia. A suitable electrolyte solution (Ringer lactate, Baxter) was administered via intravenous infusion during surgery. An anti-inflammatory drug (carprofen, Rimadyl®, Zoetis, subcutaneous) and a prophylactic antibiotic treatment (amoxicillin, Duphamox LA®, Zoetis, intramuscular) were administered via pre-operative injection. The surgical areas were clipped free of wool, scrubbed with povidone iodine (Vetedine savon®, Vetoquinol), wiped with 70% isopropyl alcohol (Savetis), painted with povidone iodine solution (Vetedine solution®, Vetoquinol), and draped. The sheep were placed in the supine position on a warmed pad. A rectal temperature probe and a rumen tube were inserted during surgery. Electrocardiogram (ECG), peripheral non-invasive arterial blood pressure, and oxygen saturation were monitored.

**Surgical procedure**

The surgery was performed in a dedicated operating theatre by a veterinary surgical specialist from Namsa, using standard aseptic techniques (Fig. 2). A cutaneous incision was made on the medial side of each femoral condyle. The muscles were separated using blunt dissection to access the femur, and the periosteum was carefully removed from the femoral epiphysis to expose the implant sites. For each site, four-step drilling sequences were conducted perpendicular to the bone surface to obtain a final hole of 3.5 mm diameter with an approximate depth of 10 mm. Each drilling step was performed at a maximum drilling speed of 1 200 rpm and followed by extensive rinsing with saline to control temperature increase at the implantation site and to remove bone debris. Following drilling, the implant was inserted with a maximum torque of 45 N.cm. The incision was closed by separately suturing the capsule, muscles, and the subcutaneous layer with absorbable thread (Ethicon® PDS™ II 1 and Ethicon® Coated Vicryl™ 2-0). The skin layer was closed with surgical staples (Appose™ ULC AutoSuture™, Covidien™). The wounds were disinfected using an iodine solution. The legs were not restrained in any manner post-surgery.

**Post-operative and terminal procedures**

The animals were observed daily for general health and to detect mortality and morbidity. The implantation sites were examined daily for adverse reactions until removal of sutures. When any animal exhibited adverse clinical signs, it was examined and treated as needed. An intramuscular injection of buprenorphine (Buprecare®, Axience) was administered at the end of the surgery day, then daily for two days post-surgery. An anti-inflammatory drug (carprofen, Rimadyl®, Zoetis) was subcutaneously injected daily for five days post-surgery and an antibiotic (amoxicillin, Duphamox LA®, Zoetis) was intramuscularly injected every two days for eight days post-surgery. The wounds were disinfected with iodine solution (Vetedine solution®, Vetoquinol) daily until two days after removal of surgical staples. The surgical staples were removed after complete healing (2 weeks following surgery). On week 4 and week 13, the designated animals were weighed and euthanized by intravenous injection of a lethal solution (Doléthal®, Vetoquinol). These intervals were chosen to evaluate local tissue effects and bone-healing performance after both a short and a midterm implantation, as suggested in ISO 10993 standard - part 6, for non-degradable materials. The distal femur was harvested and explants were fixed in 10% Neutral Buffered Formalin (NBF) for histopathologic analysis. If not used for replacement, reserve implant sites were harvested in the same way, fixed, dehydrated, embedded, and stored for potential use in subsequent analyses. The newly formed calcified mineralization fronts were marked by fluorescence via prior subcutaneous injections in the neck and the back of three fluorochromes: xylenol orange (XO), calcein green (CG), oxytetracyclin (OTC). These fluorochromes bind to calcium at sites of bone mineralization and, if injected at specific times, enable mineralization front demarcations to be distinguished. For the 4-week group, the three fluorochromes were respectively injected on day 5, day 15, and day 25. Injections for the 13-week group were performed in respectively week 4, week 8, and week 12.

## 2.6  Micro-Computed Tomography (Micro-CT)

**Micro-CT preparation**

After fixation in 10% NBF (VWR), a total of forty implanted sites, one non-implanted REF article and one non-implanted TEST article were scanned by cone beam micro-computed tomography ($\mu$CT 40, SCANCO, Switzerland). The specimens were placed in cylindrical holders to obtain transverse tomograms of the implanted article at peak kilovoltage of 70 kVp, an intensity of 114 µA, a resolution of 15 µm and an integration time of 900 ms. The implant and bone were separated from the background through segmentation performed in conjunction with a Gaussian bandpass filter, respectively applying lower and upper density thresholds of 330 mg/cm$^3$ and 2275 mg/cm$^3$ for bone and titanium.

**Micro-CT evaluation**

Two Volumes Of Interest (VOI) were defined (Fig. 3a). Each VOI consisted in a conical tube with the inner edge along the core of the article (excluding the core but including the threads) and the outer edge placed at a fixed distance of 1 mm from the inner edge. The upper VOI (VOI C) was placed in the coronal area of the article. The lower VOI (VOI A) was placed along the apical and conical remaining area of the article. For each VOI, Bone Volume (BV), Bone Volume/Total Volume of the VOI (BV/TV), Bone-to-Implant Contact (BIC), trabecular thickness (Tb.Th) and trabecular spacing (Tb.Sp) were computed. For each femoral condyle, Tb.Th and Tb.Sp were also computed for a volume of 4x4x4 mm$^3$ of trabecular healthy bone, as a reference. The non-implanted scanned implants served as a baseline to help distinguish the article from the bone.

## 2.7  Histopathology

**Histologic preparation**

After fixation in 10% NBF (VWR), a total of forty implanted sites, one non-implanted REF article and one non-implanted TEST article were dehydrated in alcohol solutions of increasing concentration, cleared in xylene and embedded in polymethylmetacrylate (PMMA). For each explant and non-implanted article, one central longitudinal section was obtained using a microcutting and grinding technique (Exakt$^{TM}$, approximately 40 $\mu$m thick), for a total of ten sections per group. The sections were left unstained for epifluorescence analysis and then stained with modified Paragon for qualitative and semi-quantitative histopathologic analyses as well as quantitative histomorphometric analysis.

**Histopathologic evaluation**

Prior to section staining, an epifluorescence analysis was conducted. Bone mineralization rate was assessed by evaluating the progression of the mineralization front at the time of the corresponding fluorochrome injection. Qualitative and semi-quantitative histopathologic evaluations of the local tissue effects at the implantation sites were conducted by an independent senior histopathologist from Namsa using a microscope (Nikon Eclipse E600, Nikon, France) coupled with a digital camera (DN 100, Nikon, France) at magnifications of x2, x4, x10, x20 and x40. Tissue damage, cellular inflammatory response, repair phase of inflammation, fatty infiltrate, and other parameters such as hemorrhage, cell degeneration, bone ingrowth, encapsulation, and bone healing were evaluated semi-quantitatively and graded using a scoring method suggested in the standard ISO 10993 – part 6. Scoring is described in Table 1.

**Histomorphometric evaluation**

As with histopathologic evaluation, histomorphometric analysis was conducted by an independent senior histopathologist Namsa, on forty modified Paragon sections. Sections were scanned (Zeiss AXIOSCAN Z1) and analyzed at a magnification of x20, with a color image analyzing system (Tribvn, France, CALOPIX 3.2.0) to perform a semi-automatic analysis. Four standardized Regions Of Interest (ROI C1, C2, A1, A2) were defined for each longitudinal section (Fig. 3b). Each ROI consisted of a rectangle with the inner edge along the core of the article (excluding the core but including the threads) and the outer edge placed at a fixed distance of 1 mm from the inner edge. The upper ROI (ROI C) was placed in the coronal area of the article. The lower ROI (ROI A) was placed along the apical and remaining conical area of the implant. Each zonal ROI (sum of ROI C1 and C2 and sum of ROI A1 and A2) of the TEST was compared to each corresponding ROI of the REF.

## 2.8 Statistical analyses

Minitab 19 was used to conduct statistical analyses. Student's t-tests were used for comparisons of the two implants, considering a *P*-value below 0.05 ($\alpha$ =0.05) as statistically significant.

# 3 Results

## 3.1 Implant surface characterization

**Scanning electron microscopy**

SEM micrographs of the implant's surface revealed significant differences between the samples' topographies. While the REF surface only shows a heterogeneous microtopography (Fig. 4a and 4b), the TEST surface topography is entirely replaced by a homogeneous nodule-like microstructure (Fig. 4c). Furthermore, the nanotopography is visible as a spike-like structure (Fig. 4d).

**Energy-dispersive X-Ray spectrometry**

The EDX spectra revealed that both REF and TEST surfaces are mostly composed of oxygen, titanium and carbon (Table 2). A low concentration of aluminum was measured on the REF surface but none was detected on the TEST surface.

**Roughness measurements**

Roughness measurements revealed few differences between REF and TEST surfaces (Table 2). While Sa and Sq parameter values were similar from one surface to another, significant differences were observed for the Sz and Sdr parameters.

**Wettability**

Wettability measurements on the TEST surface showed significantly lower contact angles for every liquid compared to the REF surface (Table 2). Therefore, the TEST surface is more hydrophilic than the REF one, showing a contact angle with water of 8.76±3.20° while 46.51±4.44° for the REF surface. While surface free energy results tend to corroborate this behavior, statistical significance was not reached ($P$=0.097).

## 3.2 Micro-CT analysis

When REF and TEST implants were compared by Micro-CT analysis, no statistical difference was found in BIC and BV/TV (Table 3). At 4 weeks after implantation, BIC reached 65.1±10.9% for the REF implant in the coronal region, against 61.8±10.8% for the TEST one. At 13 weeks, BIC increased to 74.4±10.6% for the REF implant and significantly increased to 80.6±7.4% for the TEST one. There was no evidence of variation in BV/TV ratio over time. Bone response in the apical region followed a significantly different pattern from that of the coronal region for the same period. However, neither of these parameters' values in the apical region differed significantly between implants and between time periods. While no significant differences in Tb.Th were found between the REF and TEST implants at 4 weeks, both in the coronal and in the apical regions, at 13 weeks after implantation there were significant differences between the implants in the apical region, reaching respectively 0.26±0.03 mm and 0.32±0.07 mm. A similar trend was observed in the coronal region, with respective values of 0.32±0.03 mm and 0.37±0.06 mm, although without statistical significance ($P$ = 0.075). In contrast, no differences in Tb.Sp were observed between regions or time periods.

## 3.3 Histopathologic analysis

Owing to an off-axis section, one REF implant was replaced by a reserve implant for histology. At 4 weeks (Fig. 5), qualitative analyses of explants did not reveal significant differences in impact on local tissue between the REF and the TEST implants. The defect generated through drilling during the surgical procedure was still visible for both implants, allowing distinction between host bone and healing chamber, particularly in the apical region of the implant. No marked inflammation was observed by the histopathologist, with only a small number of macrophages and osteoclasts admixed with a few lymphocytes and polymorphonuclear cells. Bone marrow formation was observed in 4 out of 10 sites for the REF implant against 7 out of 10 sites for the TEST one. Combined with qualitative observations, these results indicate for both implants a marked woven bone neoformation, without tangible evidence of bone remodeling. Bone condensation was visible around the implants, especially in the coronal region. Signs of bone apposition and moderate to marked osteoconduction were observed, along with the marked presence of osteoblasts. In particular, TEST implants quickly formed a thin and continuous film of bony tissue at their surface (Fig. 5e), as highlighted by epifluorescent analysis (Fig. 5f). Furthermore, the latter analysis showed that for both implants, bone mineralization activity had already started at day 5, reaching a peak at day 15 mostly through bone deposition, then slightly decreased at day 25 (Table 1). At 13 weeks (Fig. 6), while no significant differences were noted in the semi-quantitative bone healing parameters between the REF implant and the TEST one, differences were observed in bone response in terms of architecture. Both implants showed fewer osteoblasts than the 4W group, as well as marked signs of bone neoformation, osseointegration, and osteoconduction. Consistent bone marrow formation was found and epifluorescent analysis (Fig. 6c, f) revealed continuing strong bone mineralization activity, mostly through bone thickening and surface apposition, at week 4 and week 8, which then slowed to a more moderate level at week 12 (Table 1). Nevertheless, the TEST implant showed no more bone condensation and with thick bone trabeculae (Fig. 6e, f). The defect margins were undistinguishable between host bone and neoformed bone. For the REF implant, the defect margins remained somewhat visible and discrete bone condensation could still be observed (Fig. 6b), despite similar bone remodeling rate compared to the TEST implant (Table 1; Fig. 6c). Semi-quantitative histopathologic results at week 4 and week 13 are compiled in Table 1.

### 3.4 Histomorphometric analysis

Contrary to the Micro-CT evaluation, significant differences were pointed out by histomorphometric analyses. Only slight variations in BIC in the apical region were visible, increasing from 63.5±9.9% to 66.2±14.7% for the REF implant and decreasing from 69.2±10.3% to 64.2±14.6% for the TEST implant between week 4 and week 13 (Fig. 7). In the coronal region, the BIC of the REF implant increased from 68.3±8.8% to 74.8±13% between week 4 and week 13. BIC values for the TEST implant were significantly higher than for the REF one at both week 4 and week 13, respectively 79.3±11.2% and 86.4±6.7%. No implant impact on local tissue densities was observed. A significant decrease in fibrous tissue density was observed between week 4 and week 13 for both implants, and in both their coronal and apical regions. In particular, in the apical region this decrease led to significantly higher fibrous tissue density for the TEST implant at week 13.

## 4 Discussion

This study provides evidence that implant surface modifications have a significant impact on bone response in the coronal region. Several previous studies demonstrated that surface chemical composition modification may alter peri-implant bone response (26). In the present study, both implant surfaces, composed of titanium, seem to have adsorbed hydrocarbons from the exposure to ambient environment, explaining the presence of carbon. Moreover, traces of aluminum contamination, supposedly due to the corundum grit-blasting, were detected on the REF surface. Such pollutants may alter the nature of bone response at the peri-implant interface. Moreover, our roughness measurements point to the influence of surface processing on implant topology. Indeed, the combination of grit-blasting and dual chemical etching resulted in a significantly lower peak-to-valley height and developed surface interfacial area ratio, although the arithmetic mean deviation and root-mean-square deviation of the surface remained unaffected. It is assumed that while grit-blasting is responsible for topology modifications at the micrometric scale, chemical attacks act at lower scales, resulting in less marked differences in micro-roughness parameters. This is in fact corroborated by our wettability results. Depending on both the topography and chemical composition of the surface, the TEST contact angles for every liquid tested were significantly lower, therefore showing a more hydrophilic surface. Overall, these results, together with the histomorphometric BIC analysis, confirm previous conclusions in the scientific literature (7,14): surfaces incorporating features at both microscopic and nanoscopic scales generate an additive effect on the adhesion, proliferation, and differentiation of osteogenic cell lines, leading to increased de novo bone formation in vivo. While the exact mechanisms of overall bone response enhancement remain unclear, studies suggest that the higher surface energy resulting from incorporation of nanofeatures positively affects the adsorption and conformation of vitronectin and fibronectin, proteins responsible for osteoblast adhesion, onto the implant surface. Such adhesion appears to regulate the subsequent proliferation and differentiation of mesenchymal stem cells (27,28).

The absence of tissue deleterious response for both REF and TEST implants indicates that the implants' biocompatibility was preserved despite the topography and chemical modifications in this study. Furthermore, mineralization activity was not affected by the TEST surface treatment. In a study conducted in dog jaws, Abrahamsson et al. (2004) suggested that there was an increase in lamellar bone formation around dental implants between 2 weeks and 4 weeks after implantation (29), which is similar to the peaks observed here for the REF and TEST implants 15 days after implantation. This same study also emphasized the presence of a front of bone deposition onto the implant surface. The semi-quantitative histopathologic evaluation used here suggests that the TEST surface treatment was not deleterious to bone mineralization activity and moreover promoted osteoconduction. Micro-CT evaluation demonstrated no significant difference between the REF and TEST implants regarding BIC and BV/TV parameters, in both coronal and apical regions. This lack of difference compared to the findings from histomorphometric analyses could be explained by the high interferences generated by titanium implants (30). These artifacts hinder measurements close to the surface of implants, thus decreasing the accuracy of such evaluation. Schwartz et al. (2008) precluded assessment by Micro-CT of bone healing at the interface due to these artifacts, as well as to poor acquisition resolution (9). However, our analysis highlighted different implant osseointegration behaviors. First, while Tb.Th in the apical region did not significantly differ according to implant at 4 weeks, it was significantly higher for the TEST implant at 13 weeks. A similar trend, although without statistical significance, was observed in the coronal region, which gives additional weight to this observation. It should be noted that major differences in bone healing were observed within a given group of implants between the coronal and the apical regions. This could be explained by a different implant thread design in the apical region, altering local stress distribution at the interface with bone (31,32) and increasing wound chamber dimensions (33), thereby modifying the osteogenic response to the implant (34).

These observations need to be viewed in conjunction with the histopathologic findings. Indeed, while at 4 weeks REF and TEST implants induced similar bone architecture, with bone condensation and presence of the initial defect margins, at 13 weeks the TEST implant induced bone healing, with trabeculae recovering their connectivity and no detectable trace of the drilling for implant insertion. In addition, trabeculae appeared to be oriented towards the thread tips of the implant, where interfacial mechanical stress is significant. This is in accordance with Wolff's law of bone remodeling (35), which led to the concept of "bone functional adaptation" (36), explaining the process of long-term osseointegration. According to this concept, mechanical load applied to living bone influences the structure of bone tissue over time. Increased local strain results in greater deposition of bone tissue while decreased strain leads to resorption of bone tissue until the original bone strain levels, also termed "optimum customary strain level", are restored Interestingly, similar patterns were observed in other in vivo studies. Perrin et al. (2002) inserted titanium dental implants of various topographies into Land Race pig mandibles, observing a preferential orientation and distribution of bone trabeculae after 10 weeks. These trabeculae were oriented perpendicular to the implant surface and located at the thread level of the implant (37). The authors explained this behavior with reference to Gross et al. (1990), who inserted titanium cylinders of differing surface roughness into the distal epiphysis of rabbit femurs and noted the formation of a cortical shell around implants with smooth surfaces, opposed to oriented trabeculae with increased surface roughness (38). It was suggested that surface roughness promotes bone trabecularization around the implant, leading to better immobilization. In an animal experiment employing porous coated and proximally partially porous coated femoral canine implants, Bobyn et al. (1987) noted a similar shell around the smooth surface, interposed with a space filled with fibrous tissue (39). Based on these observations, Luo et al. (1999) used computational methods to indicate that such a shell should resorb over time, leading to trabecularization of the bone-implant interface (40).

Clinically, *in vivo* investigations report successful osseointegration of dental implants for BIC between 50% and 80% (41). Considering the current experimental conditions and results, it is therefore assumed that both REF and TEST implants would achieve long-term stability. However, by positively affecting coronal BIC results both at 4 weeks and 13 weeks compared to the REF implant, it is supposed by the authors that the surface treatment applied to the TEST implant may be of interest for improving osseointegration of implants in bone of poor quality such as osteoporotic bone (19).

Bone healing and remodeling mechanisms are complex, and this study faced difficult choices. On the choice of animal model, canine and porcine models are known to provide bone healing and remodeling rates closer to humans than the sheep model. However, the sheep was shown to be a relevant model in many intraosseous implantation studies published in scientific peer-reviewed journals and is recognized by International Regulatory Organisms (42,43). Compared to small animal models, the sheep model offers the advantage of a body weight, as well as bone healing and remodeling patterns, more similar to the human (44). Furthermore, this model currently raises fewer ethical issues than models such as the dog, and is easier to handle, more amenable to intervention, than the porcine model (44). On the choice of implantation site, it is universally recognized that the intended anatomical location of dental implants, alveolar bone from the jaw, is not equivalent to femoral bone in its origin, method of ossification, microstructure or rate of turnover. In addition, obtaining data related to dental implant evaluation, such as buccal and lingual bone crest levels, appears impractical from the femur, which partially limits the range of analyses specifically related to dental implants. Nonetheless, the sheep femoral implantation model is highly standardized and associated with low rates of complication. The osseointegration properties of dental implants can be assessed with minimal variability, enabling finer comparisons of surface treatment influence than with jawbone implantation (45). Histopathological analyses were performed by only one senior histopathologist who provided only one value per parameter and per sample. Therefore, unfortunately, the reproducibility of results for the same sample at different times cannot be clearly supported by any reliability data. Furthermore, implant dimensions constrained the sampling to only one longitudinal section per explant. However, the authors are confident that the preparation of ten slices per group, along with the assessment of the overall peri-implant bone architecture through micro-CT analyses, reduced any potential bias generated by the study design.

# 5 Conclusion

The present pre-clinical study evaluated the biocompatibility and osseointegration of dental implants with different surface characteristics in sheep femoral condyles. The REF implant surface was obtained by grit-blasting while the TEST implant surface resulted from a succession of grit-blasting, acid-etching, and basic-etching, the two latter inducing nanofeatures at its surface. After 4 and 13 weeks of implantation, both implants showed good biocompatibility within their environment, without any deleterious local effect on the surrounding tissue. Bone apposition was significantly higher at both 4 and 13 weeks for the TEST implant, with thicker bone trabeculae in the apical region of the implant. This suggests that the presence of nanofeatures at the surface of TEST dental implants promoted faster bone healing. Finally, the distinction between the coronal and apical regions of the implant demonstrates that bone behavior differed along the implant, possibly due to differing implant thread geometry and the greater local stresses generated at the apex.

# Figure legends

Figure 1: (a) Full-Size High-Resolution SEM image of REF and TEST implant used for the study; (b) General implant dimensions.

Figure 2: Surgery procedure for implantation. (a) Drilling at 1 200 rpm; (b) dental implant insertion at 45 N.cm; (c,d) implant positions on femoral condyle after completed surgery. Three sites per condyle are allocated for implants.

Figure 3: (a) Tomogram obtained by Micro-CT tomography, presenting the different Volumes Of Interest (VOI A and C); (b) Representative photomicrograph used for histopathologic analysis, introducing four Regions Of Interest (ROI A1, A2, C1 and C2).

Figure 4: SEM of implant surface (a,b) REF surface ; (c,d) TEST surface. The REF surface shows a heterogeneous microtopography while the TEST surface has a homogeneous nodule-like microstructure and a spike-like nanotopography.

Figure 5: Photomicrograph of a (a,b,c) REF implant and (d,e,f) TEST implant, 4 weeks after implantation. Bone condensation is visible all around the implant, depicting the initial defect margins created through the surgical drilling. Signs of osseointegration are visible through woven bone neoformation at the surface of both REF and TEST implants. Under epifluorescence microscopy (c,f), calcein green (15 days) revealed bone mineralization at the surface of the TEST implant, where a continuous thin film of bony tissue is formed, while discrete and discontinuous on the REF implant (white lines). OB: old bone; OI: osseointegration; BM: bone marrow; NFB: Newly formed bone.

Figure 6: Photomicrograph of (a,b,c) REF implant and (d,e,f) TEST implant, 13 weeks after implantation. While defect margins are still visible (black arrows) along with discrete bone condensation around the REF implant, the TEST implant showed almost invisible bone condensation and defect margins. Under epifluorescence microscopy (c,f), bone thickening and surface apposition are still observable for both implants, in xylenol orange (4 weeks) and calcein green (8 weeks). Oxytetracyclin (13 weeks) revealed slowing bone mineralization. OB: old bone; OI: osseointegration; BM: bone marrow; NFB: Newly formed bone; DM: drill margins.

Figure 7: Histomorphometric analysis of BIC (%) in the coronal and apical portions of REF and TEST implants. * $P<0.05$ compared to REF group at the same time period.